\documentclass{aastex701}

\usepackage{xspace}

\begin{document}

\title{TSSC comet-centered data products from TESS 3I/ATLAS observations}

\author[orcid=0000-0002-7395-4935]{Jorge Martinez-Palomera}
\affiliation{University of Maryland, Baltimore County, MD, USA}
\affiliation{NASA Goddard Space Flight Center, MD, USA}
\email[show]{jorge.i.martinezpalomera@nasa.gov}  

\author[orcid=0000-0002-2830-9064]{Amy Tuson}
\affiliation{University of Maryland, Baltimore County, MD, USA}
\affiliation{NASA Goddard Space Flight Center, MD, USA}
\email{}

\author{TESS Science Support Center}
\affiliation{NASA Goddard Space Flight Center, MD, USA}
\email{}

\begin{abstract}

3I/ATLAS is the third known interstellar object to pass through our Solar System. NASA's Transiting Exoplanet Survey Satellite (TESS) made dedicated observations of 3I/ATLAS between 15 -- 22 January 2026 (Sector 1751), capturing high-cadence observations at 200s and 20s cadence. We present two High Level Science Products (HLSPs): (1) comet-centered image time series, corrected for background scattered light and stars; and (2) aperture light curves extracted from the corrected images. We created these data products using the official TESS products and they are publicly available at the Mikulski Archive for Space Telescopes (MAST). TESS's high-precision, near-continuous photometry will provide unique insights into the comet's activity following its closest approach to the Sun. The TESS Science Support Center (TSSC) has created these data products to facilitate scientific analyses by the TESS and Solar System communities.

\end{abstract}

\keywords{\uat{Interstellar objects}{52} --- \uat{Comets}{280} --- \uat{Time domain astronomy}{2109} --- \uat{Photometry}{1234}}

\section{Introduction}\label{sec:intro}

The interstellar comet 3I/ATLAS was discovered on 1 July 2025 by the ATLAS survey \citep{2025ApJ...989L..36S}. It has orbital eccentricity of $\sim\!6.1$, perihelion of $\sim\!1.36$\,au, inclination of $\sim\!175^\circ$ and velocity at infinity of $\sim\!57\text{km\,s}^{-1}$ \citep{2025MNRAS.542L.139B}. 3I/ATLAS has been extensively monitored by ground- and space- based observatories in order to characterize its unique behavior and obtain more insights into the population of interstellar comets.

TESS \citep{2015JATIS...1a4003R} serendipitously observed 3I/ATLAS between 7 May and 2 June 2025 \citep{2025ApJ...991L...2F,2025ApJ...994L..51M}. 
Between 15 -- 22 January 2026, TESS conducted dedicated observations of 3I/ATLAS. This pointing was designated sector 1751 \citep{https://doi.org/10.17909/mrs3-bw92} and consisted of 3 blocks of data: 18hr on 15 January 2026, 10hr on 18--19 January 2026, and 73hr on 19--22 January 2026. The two data gaps were due to contingency mode and data downlink, respectively. The observations placed the comet on TESS camera 2 CCD 3 for the entire sector\footnote{See the \href{https://heasarc.gsfc.nasa.gov/docs/tess/tess-special-news-bulletin-dec-18th.html}{TSSC webpage} for details.}. TESS collected full frame images (FFIs) with a 200s cadence, and target pixel files (TPFs) with 20s and 120s cadences. 

Here we present the creation of two data products using the Science Processing Operations Center \citep[SPOC;][]{spoc} FFIs: (1) comet-centered, background-corrected image time series, and (2) light curves extracted from these images. These data products follow a similar format to the mission products (TPFs and light curve files). We used the \texttt{tess-asteroids} Python package \citep{tuson_2025_17662309} to extract and process the data. This package enables efficient retrieval of FFI data, background correction, and photometric extraction for moving objects observed by TESS.

\section{Comet-centered Image Time Series}\label{sec:mtpf}

We used \texttt{tess-asteroids v1.5.0}, with minor modifications to optimize for a comet with an extended tail, to create an image cutout centered on 3I/ATLAS. Our data extraction and processing followed these steps:

\begin{itemize}
    \item \textbf{Ephemeris:} we used JPL Horizons to obtain the ephemeris of 3I/ATLAS between 15 -- 22 January 2026. We used \texttt{tesswcs}\footnote{\href{https://github.com/tessgi/tesswcs}{\texttt{tesswcs}} is software to retrieve TESS world coordinate system solutions for every camera and CCD.} to convert sky coordinates into pixel coordinates on the TESS CCD.
    
    \item \textbf{Pixel time series:} we used \texttt{tesscube}\footnote{\href{https://github.com/tessgi/tesscube}{\texttt{tesscube}} is software to access TESS FFIs on MAST cloud storage environment.} to access the calibrated FFIs. We created a $111\times71$ pixel cutout centered on the predicted position of the comet in all 1835 available frames. The cutout size was selected to ensure the entire comet tail was enclosed in the image.

    \item \textbf{Time correction:} we used \texttt{lkspacecraft}\footnote{\href{https://github.com/lightkurve/lkspacecraft}{\texttt{lkspacecraft}} is software that uses SPICE kernels to compute TESS spacecraft state vectors.} to compute the barycentric time correction at the position of the target in each frame. We present the time stamps in Barycentric Dynamical Time.

    \item \textbf{Background correction:} we created a background model for the image cutouts consisting of two components:
    \begin{enumerate}

    \item \textbf{Background scattered light:} we modeled this signal at each frame with a space-dependent linear model that uses PCA components to interpolate the flux over stars. We masked out pixels with background stars and the comet from the modeling using $\sigma$-clipping statistics and the comet's ephemeris, respectively. This creates a scattered light model for all pixels at each time in the data.
    
    \item \textbf{Background stars:} we model the stars in the field with a time-dependent linear model to allow for variability. We model the pixel light curves in a 1-day time window centered on the current frame; this process is iterated for each pixel and frame. Pixels with comet signal are masked out from the modeling. This models every pixel that contains star flux.

    \end{enumerate}
    
    We subtracted the background scattered light and star model from the flux. 
    
    The background correction was an iterative process. This was necessary to refine the target mask and ensure the comet's nucleus and tail were fully covered.

    \item \textbf{Pixel quality mask:} we created a pixel mask at each time to flag anomalies, e.g. saturated pixels, strap columns and pixels with a poorly constrained background model.
\end{itemize}

\begin{figure*}
\begin{interactive}{animation}{fig1anim.mp4}
\includegraphics[width=1\textwidth]{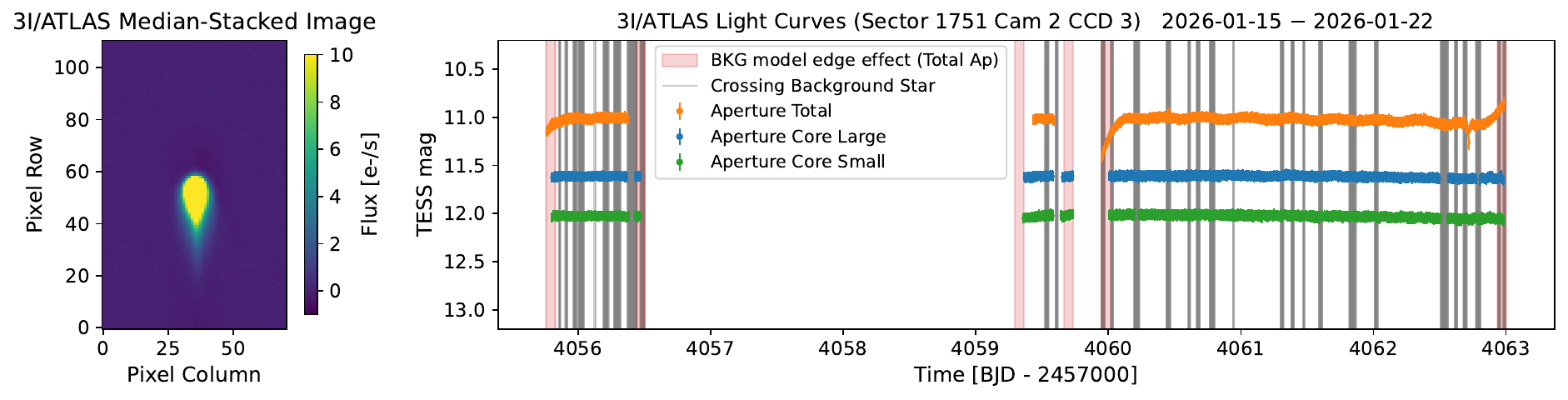}
\end{interactive}
\caption{\textit{Left:} median-stack image generated from the
background-corrected data and frames with SPOC quality flag zero.
\textit{Right:} extracted light curves from a core small aperture
(green), core large aperture (blue), and total aperture (orange).
We highlight frames where the comet passes over background stars
(grey) and frames with background model edge effect caused by the
comet's tail (red).\\
An animation of the comet-centered and background-corrected image time series is available in the online Journal.
\label{fig1}}
\end{figure*}

The left panel of Figure \ref{fig1} shows the median-stack of all frames with SPOC quality flag 0. The image shows that 3I/ATLAS has a prominent core and extended tail.

\section{Light Curves}\label{sec:lcf}

We extracted three light curves from our background-corrected images. As 3I/ATLAS shows an extended tail and core signal, we defined three apertures:

\begin{itemize}
    \item \textbf{Core apertures:} we used \texttt{lkprf}\footnote{\href{https://github.com/lightkurve/lkprf}{\texttt{lkprf}} is software to evaluate the TESS PRF model at any position on the detectors.} to evaluate the TESS pixel response function (PRF) model at the comet's position in each frame. We define a \textbf{core small} and a \textbf{core large} aperture from pixels with $\geq1\%$ and $\geq0.1\%$ of the PRF's total flux, respectively. In this way we cover the inner and extended nucleus, separately.
    
    \item \textbf{Total aperture:} we defined an aperture that covers the nucleus and extended tail by selecting pixels with a flux value $\geq1\text{e}^-\text{s}^{-1}$ in the median-stacked image (see Figure \ref{fig1}).
    
\end{itemize}

In each case, we converted the measured flux into a TESS magnitude using a zero-point of $20.44 \pm 0.05$\footnote{\url{https://archive.stsci.edu/missions/tess/doc/TESS_Instrument_Handbook_v0.1.pdf}}. The median TESS magnitudes from our light curves are $T_{\text{core\_s}} = 12.03 \pm 0.01$, $T_{\text{core\_l}} = 11.61 \pm 0.01$, and $T_{\text{total}}=11.02 \pm 0.01$. The three extracted light curves are shown in the right panel of Figure \ref{fig1}. We computed several metrics to assess the quality of the photometric measurements at the cadence and pixel level, such as fraction of good quality pixels in the aperture, background light curves, and background model statistics. These metrics are available for both image and light curve files.

\section{Data Availability}\label{sec:data}

All our data products, stored as FITS files, are available at MAST as a High-Level Science product (HLSP): \dataset[10.17909/zefb-7f89]{https://doi.org/10.17909/zefb-7f89}
and in a Zenodo repository: \dataset[10.5281/zenodo.18489884]{https://doi.org/10.5281/zenodo.18489884} \citep{martinez_palomera_2026_18489884}. 
Additionally, we published a \href{https://github.com/tessgi/tess-3i}{GitHub} repository\footnote{\url{https://github.com/tessgi/tess-3i}} 
with Jupyter notebooks that explain the data extraction and processing, and tutorials on how to open the data products.
We will create similar data products using the 20s and 120s cadence data when they become available at MAST.

\begin{acknowledgments}

This paper includes data collected with the TESS mission, obtained from the MAST data archive at the Space Telescope Science Institute (STScI). Funding for the TESS mission is provided by the NASA Explorer Program.
Funding for this work for JMP and AT is provided by NASA grant 80NSSC20M0192. The material is based upon work supported by NASA under award number 80GSFC24M0006.
\end{acknowledgments}
\bibliography{new.ms}{}

@misc{https://doi.org/10.17909/mrs3-bw92,
  doi = {10.17909/MRS3-BW92},
  url = {https://archive.stsci.edu/doi/resolve/resolve.html?doi=10.17909/mrs3-bw92},
  author = {{TESS Team}},
  title = {TESS Calibrated Full Frame Images: Sector 1751},
  publisher = {STScI/MAST},
  year = {2026}
}

@ARTICLE{2015JATIS...1a4003R,
       author = {{Ricker}, George R. and {Winn}, Joshua N. and {Vanderspek}, Roland and {Latham}, David W. and {Bakos}, G{\'a}sp{\'a}r {\'A}. and {Bean}, Jacob L. and {Berta-Thompson}, Zachory K. and {Brown}, Timothy M. and {Buchhave}, Lars and {Butler}, Nathaniel R. and {Butler}, R. Paul and {Chaplin}, William J. and {Charbonneau}, David and {Christensen-Dalsgaard}, J{\o}rgen and {Clampin}, Mark and {Deming}, Drake and {Doty}, John and {De Lee}, Nathan and {Dressing}, Courtney and {Dunham}, Edward W. and {Endl}, Michael and {Fressin}, Francois and {Ge}, Jian and {Henning}, Thomas and {Holman}, Matthew J. and {Howard}, Andrew W. and {Ida}, Shigeru and {Jenkins}, Jon M. and {Jernigan}, Garrett and {Johnson}, John Asher and {Kaltenegger}, Lisa and {Kawai}, Nobuyuki and {Kjeldsen}, Hans and {Laughlin}, Gregory and {Levine}, Alan M. and {Lin}, Douglas and {Lissauer}, Jack J. and {MacQueen}, Phillip and {Marcy}, Geoffrey and {McCullough}, Peter R. and {Morton}, Timothy D. and {Narita}, Norio and {Paegert}, Martin and {Palle}, Enric and {Pepe}, Francesco and {Pepper}, Joshua and {Quirrenbach}, Andreas and {Rinehart}, Stephen A. and {Sasselov}, Dimitar and {Sato}, Bun'ei and {Seager}, Sara and {Sozzetti}, Alessandro and {Stassun}, Keivan G. and {Sullivan}, Peter and {Szentgyorgyi}, Andrew and {Torres}, Guillermo and {Udry}, Stephane and {Villasenor}, Joel},
        title = "{Transiting Exoplanet Survey Satellite (TESS)}",
      journal = {Journal of Astronomical Telescopes, Instruments, and Systems},
         year = 2015,
        month = jan,
       volume = {1},
          eid = {014003},
        pages = {014003},
          doi = {10.1117/1.JATIS.1.1.014003},
       adsurl = {https://ui.adsabs.harvard.edu/abs/2015JATIS...1a4003R}
}

@ARTICLE{2025ApJ...994L..51M,
       author = {{Martinez-Palomera}, Jorge and {Tuson}, Amy and {Hedges}, Christina and {Dotson}, Jessie and {Barclay}, Thomas and {Powell}, Brian},
        title = "{Prediscovery TESS Observations of Interstellar Object 3I/ATLAS}",
      journal = {\apjl},
         year = 2025,
        month = dec,
       volume = {994},
       number = {2},
          eid = {L51},
        pages = {L51},
          doi = {10.3847/2041-8213/ae1f91},
archivePrefix = {arXiv},
       eprint = {2508.02499},
 primaryClass = {astro-ph.EP},
       adsurl = {https://ui.adsabs.harvard.edu/abs/2025ApJ...994L..51M}
}

@ARTICLE{2025ApJ...991L...2F,
       author = {{Feinstein}, Adina D. and {Noonan}, John W. and {Seligman}, Darryl Z.},
        title = "{Precovery Observations of 3I/ATLAS from TESS Suggest Possible Distant Activity}",
      journal = {\apjl},
         year = 2025,
        month = sep,
       volume = {991},
       number = {1},
          eid = {L2},
        pages = {L2},
          doi = {10.3847/2041-8213/adfd4d},
archivePrefix = {arXiv},
       eprint = {2507.21967},
 primaryClass = {astro-ph.EP},
       adsurl = {https://ui.adsabs.harvard.edu/abs/2025ApJ...991L...2F}
}

@ARTICLE{2025MNRAS.542L.139B,
       author = {{Bolin}, Bryce T. and {Belyakov}, Matthew and {Fremling}, Christoffer and {Graham}, Matthew J. and {Abdelaziz}, Ahmed M. and {Elhosseiny}, Eslam and {Gray}, Candace L. and {Ingebretsen}, Carl and {Jewett}, Gracyn and {Lisse}, Carey M. and {Karpov}, Sergey and {Kilic}, Mukremin and {Ma{\v{s}}ek}, Martin and {Molham}, Mona and {Roderick}, Diana and {Takey}, Ali and {Abron}, Laura-May and {Coughlin}, Michael W. and {Hsieh}, Cheng-Han and {Noll}, Keith S. and {Wong}, Ian},
        title = "{Interstellar comet 3I/ATLAS: discovery and physical description}",
      journal = {\mnras},
         year = 2025,
        month = sep,
       volume = {542},
       number = {1},
        pages = {L139-L143},
          doi = {10.1093/mnrasl/slaf078},
archivePrefix = {arXiv},
       eprint = {2507.05252},
 primaryClass = {astro-ph.EP},
       adsurl = {https://ui.adsabs.harvard.edu/abs/2025MNRAS.542L.139B}
}

@software{tuson_2025_17662309,
  author       = {Tuson, Amy and
                  Martínez-Palomera, Jorge and
                  Hedges, Christina},
  title        = {altuson/tess-asteroids: v1.4.2},
  month        = nov,
  year         = 2025,
  publisher    = {Zenodo},
  version      = {v1.4.2},
  doi          = {10.5281/zenodo.17662309},
  url          = {https://doi.org/10.5281/zenodo.17662309},
  swhid        = {swh:1:dir:897e438d7db945c984a5f5860a79f822d4aef902
                   ;origin=https://doi.org/10.5281/zenodo.15882329;vi
                   sit=swh:1:snp:ac42a474f23496da84d35d9d1d4b53b8399f
                   141a;anchor=swh:1:rel:f56e247825c05b35692e01594789
                   5beacef796dc;path=altuson-tess-asteroids-c97aa22
                  },
}

@dataset{martinez_palomera_2026_18489884,
  author       = {Martínez-Palomera, Jorge and
                  Tuson, Amy and
                  TESS Science Support Center},
  title        = {TSSC comet-centered data products from TESS
                   3I/ATLAS observations
                  },
  month        = feb,
  year         = 2026,
  publisher    = {Zenodo},
  version      = {v2.0},
  doi          = {10.5281/zenodo.18489884},
  url          = {https://doi.org/10.5281/zenodo.18489884},
}

@ARTICLE{2025ApJ...989L..36S,
       author = {{Seligman}, Darryl Z. and {Micheli}, Marco and {Farnocchia}, Davide and {Denneau}, Larry and {Noonan}, John W. and {Hsieh}, Henry H. and {Santana-Ros}, Toni and {Tonry}, John and {Auchettl}, Katie and {Conversi}, Luca and {Devog{\`e}le}, Maxime and {Faggioli}, Laura and {Feinstein}, Adina D. and {Fenucci}, Marco and {Ferrais}, Marin and {Frincke}, Tessa and {Gillon}, Michael and {Hainaut}, Olivier R. and {Hart}, Kyle and {Hoffman}, Andrew and {Holt}, Carrie E. and {Hoogendam}, Willem B. and {Huber}, Mark E. and {Jehin}, Emmanuel and {Kareta}, Theodore and {Keane}, Jacqueline V. and {Kelley}, Michael S.~P. and {Lister}, Tim and {Mandt}, Kathleen and {Manfroid}, Jean and {Mar{\v{c}}eta}, Du{\v{s}}an and {Meech}, Karen J. and {Amine Miftah}, Mohamed and {Morgan}, Marvin and {Oca{\~n}a}, Francisco and {Pe{\~n}a-Asensio}, Eloy and {Shappee}, Benjamin J. and {Siverd}, Robert J. and {Taylor}, Aster G. and {Tucker}, Michael A. and {Wainscoat}, Richard and {Weryk}, Robert and {Wray}, James J. and {Yaginuma}, Atsuhiro and {Yang}, Bin and {Ye}, Quanzhi and {Zhang}, Qicheng},
        title = "{Discovery and Preliminary Characterization of a Third Interstellar Object: 3I/ATLAS}",
      journal = {\apjl},
         year = 2025,
        month = aug,
       volume = {989},
       number = {2},
          eid = {L36},
        pages = {L36},
          doi = {10.3847/2041-8213/adf49a},
archivePrefix = {arXiv},
       eprint = {2507.02757},
 primaryClass = {astro-ph.EP},
       adsurl = {https://ui.adsabs.harvard.edu/abs/2025ApJ...989L..36S}
}

@INPROCEEDINGS{spoc,
       author = {{Jenkins}, Jon M. and {Twicken}, Joseph D. and {McCauliff}, Sean and {Campbell}, Jennifer and {Sanderfer}, Dwight and {Lung}, David and {Mansouri-Samani}, Masoud and {Girouard}, Forrest and {Tenenbaum}, Peter and {Klaus}, Todd and {Smith}, Jeffrey C. and {Caldwell}, Douglas A. and {Chacon}, A.~D. and {Henze}, Christopher and {Heiges}, Cory and {Latham}, David W. and {Morgan}, Edward and {Swade}, Daryl and {Rinehart}, Stephen and {Vanderspek}, Roland},
        title = "{The TESS science processing operations center}",
    booktitle = {Software and Cyberinfrastructure for Astronomy IV},
         year = 2016,
       editor = {{Chiozzi}, Gianluca and {Guzman}, Juan C.},
       series = {Society of Photo-Optical Instrumentation Engineers (SPIE) Conference Series},
       volume = {9913},
        month = aug,
          eid = {99133E},
        pages = {99133E},
          doi = {10.1117/12.2233418},
       adsurl = {https://ui.adsabs.harvard.edu/abs/2016SPIE.9913E..3EJ}
}
\bibliographystyle{aasjournalv7}

\end{document}